\documentclass[prl,aps,twocolumn,showpacs,superscriptaddress,amsmath,amssymb,floatfix]{revtex4}
\usepackage{graphicx}

\newcommand{\be}{\begin{equation}}
\newcommand{\ee}{\end{equation}}

\newcommand{\rr}{{\mathbf r}}

\newcommand{\eq}[1]{(\ref{eq:#1})}
\newcommand{\eqname}[1]{\label{eq:#1}}

\begin{document}
\title{Are non-equilibrium Bose-Einstein condensates superfluid?}
\author{Michiel Wouters}
\affiliation{Institute of Theoretical Physics, Ecole Polytechnique F\'ed\'erale de Lausanne (EPFL), CH-1015 Lausanne, Switzerland}
\author{Iacopo Carusotto}
\affiliation{CNR INFM-BEC Center and Dipartimento di Fisica, Universit\`a
di Trento, I-38123 Povo, Italy}
\begin{abstract}
We theoretically study the superfluidity properties of a non-equilibrium Bose-Einstein condensate of exciton-polaritons in a semiconductor microcavity. The dynamics of the condensate is described at mean-field level in terms of a modified Complex Ginzburg Landau equation. 
A generalized Landau criterion is formulated which estimates the onset of the drag force on a small moving defect. Metastability of supercurrents in multiply connected geometries persists up to higher flow speeds.
\end{abstract}
\maketitle

Superfluidity is among the most remarkable consequences of macroscopic quantum coherence in condensed matter systems and manifests itself in a number of fascinating effects~\cite{leggett,SSLP-book}. A unified description of these phenomena is obtained in the framework of the so-called two-fluid hydrodynamics, in which the macroscopic condensate wavefunction adds up to the standard hydrodynamic variables~\cite{huang}. 
The phenomenon of macroscopic coherence is not restricted to systems at (or close to) thermodynamical equilibrium such as liquid Helium, ultracold atomic gases, or superconducting materials, but has been observed also in systems far from thermodynamical equilibrium, whose state is determined by a dynamical balance of driving and losses. Most remarkable examples are lasers and, more recently, Bose-Einstein condensates of magnons in magnetic solids~\cite{BEC-magnons} and exciton-polaritons in semiconductor microcavities~\cite{kasprzak,maxime-small}. In particular, the issue of superfluidity in this latter system has attracted a significant interest from both the theoretical~\cite{iac-superfl,vortexprobe} and experimental~\cite{amo-superfl,TOPO,vortexMadrid} points of view.

Recent experiments with resonantly pumped polariton condensates~\cite{amo-superfl} have demonstrated superfluidity as a dramatic reduction in the intensity of resonant Rayleigh scattering as originally predicted in~\cite{iac-superfl}.
The situation is less clear in the case of non-resonant~\cite{kasprzak} or parametrical (OPO)~\cite{TOPO} pumping schemes: recent experiments~\cite{TOPO} have observed propagation of polariton bullets without apparent friction, which is in contrast with the predictions of a na\"ive Landau criterion based on the elementary excitation spectrum predicted in~\cite{Goldstone}.
Another aspect of superfluidity, namely metastability of supercurrent in multiply-connected geometries was investigated theoretically in~\cite{vortexprobe} and experimentally confirmed in~\cite{vortexMadrid}.
The present paper reports a comprehensive theoretical investigation of the meaning of superfluidity for polariton condensates under a non-resonant pumping. Emphasis will be given to the novel features that originate from their non-equilibrium character.

At the mean-field level, the condensate dynamics can be described in terms of the so-called Gross-Pitaevskii equation (GPE)~\cite{SSLP-book}, which was recently generalized to non-equilibrium condensates by including the effect of pumping and losses~\cite{nonresonant,keeling}.
This mean-field description has been able to explain a number of experimental observations on polariton condensates, e.g. the ring-shaped momentum distribution of spatially narrow condensates~\cite{maxime-small,noi-shape}, the synchronization/desynchronization transition~\cite{synchro}, the spontaneous appearance of vortices~\cite{vortices}.
Nonetheless, the implicit assumption that the pumping mechanism is not frequency-selective can lead to unphysical predictions, e.g. that in a spatially homogeneous or ring-like geometry condensation is equally likely to occur in any momentum state. 
Kinetic calculations~\cite{kinetic} have pointed out the significant energy dependence of the polariton-polariton scattering processes that are responsible for replenishing the condensate. Including this feature as an energy-dependent amplification mechanism turned out to be crucial in order to extract physically meaningful predictions for the condensate fluctuations in the Wigner Monte Carlo simulations of~\cite{fluctuations}.

A simplest generalization of the GPE to include frequency-dependent pumping has the form:
\begin{multline}
	i \frac{d\psi}{dt} = \left\{ -\frac{\hbar}{2m}\nabla^2 + V_{ext} \right. \\
			\left.	+ \frac{i}{2}\left[ P\,\left(1- \frac{i}{\Omega_K}  \frac{d}{dt}\right)-r |\psi|^2
			-\gamma\right]
			        +  g |\psi|^2  \right\} \psi.
	\label{eq:CGLE}
\end{multline}
The energy zero has been set for convenience at the bottom of the lower polariton branch; the efficiency of amplification (proportional to the pumping strength $P$) decreases to zero a frequency interval $\Omega_K$ above it. Assuming a linear form of the frequency dependence of the amplification, the generalized GPE \eq{CGLE} maintains a temporally local form.
The others terms describing gain saturation ($r$), losses ($\gamma$), polariton mass ($m$), polariton-polariton interactions ($g$),  external potential  ($V_{ext}$) have the same meaning as in previous works~\cite{nonresonant,keeling}.

We first consider the evolution of the system starting from an initial state with no condensate $\psi=0$.
If the strength of pumping is enough to overcome losses $P>\gamma$, the $\psi=0$ state is dynamically unstable against the creation of a finite condensate amplitude in all the low-momentum modes for which $\hbar k^2/2m <\Omega_K\,(1-\gamma/P)$. The rate of this instability is maximum at $k=0$ and decreases for increasing $k$.
This fact is in agreement with the experimental observation of condensation naturally occurring around $k=0$ as soon as the sample is sufficiently large and free from disorder~\cite{kasprzak}.

In spite of this natural preference, condensation can be forced to occur in finite momentum state by seeding the system with a short resonant pulse at the desired $k_c$ as proposed and numerically assessed in Ref.~\cite{vortexprobe}. A related configuration was experimentally demonstrated for the OPO pump scheme in~\cite{vortexMadrid}.
The effect of the frequency-dependent pumping is then visible in the state equation relating the pumping strength $P$ to the condensate density $n_c=|\psi|^2$ in the stationary state:
\begin{equation}
|\psi|^2=\left[1-\frac{1}{\Omega_K}\left(\frac{\hbar k_c^2}{2m}+g\,|\psi|^2\right)\right]\,\frac{P}{r}.
\eqname{eqstate}
\end{equation}
As a consequence of the frequency dependent pumping, the interaction-induced blue-shift of the condensate frequency $\omega_c=k_c^2/2m+g|\psi|^2$ is responsible for a slower increase of density with pumping strength $P$. 

\begin{figure}[htpb]
\begin{center}
\includegraphics[width=0.9\columnwidth,angle=0,clip]{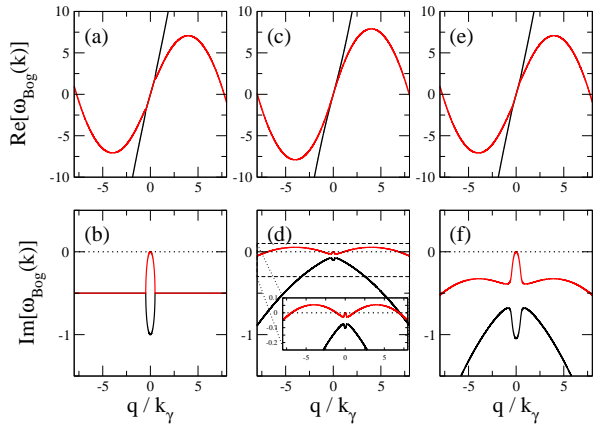}  
\end{center}
\caption{Elementary excitation spectrum of a moving, spatially homogeneous, non-equilibrium condensate as described by the linearized CGLE around the steady-state solution.
Parameters: $k_c/k_\gamma=4$, $r/\gamma=1$, $g/\gamma=1$, $\Omega_K/\gamma=50$, $n_c g/\gamma=0.1$ (c,d) and $n_c g/\gamma=1$ (e,f). Inset: magnified view of (d). Panels (a,b): same as panels (e,f) in the limit of a frequency-independent pumping $\Omega_K=\infty$. Momenta are measured in units of $k_\gamma=\sqrt{m\gamma/\hbar}$.}
\label{fig:lindisp}
\end{figure}

The dynamical stability of a moving condensate is to be assessed by linearization of the CGLE around the stationary solution. Stability requires that the imaginary part of the frequency  is negative for all Bogoliubov modes. Examples of the Bogoliubov dispersion $\omega_{Bog}(q)$ for moving condensates are shown in Fig.\ref{fig:lindisp}.
In the left-most panels (a,b), the limit $\Omega_K\rightarrow \infty$ of a negligible frequency-dependence of pumping is considered. 
Apart for the global Doppler tilting of the real part due to the finite condensate velocity, the plotted dispersion fully recovers the diffusive character first discussed in~\cite{nonresonant,keeling,Goldstone}: a flat dispersion around the condensate wavevector and an imaginary part quadratically growing in $q$. In this limit, stable condensates exist for any value of $k_c$. 
 
The effect of a frequency-selective pumping is addressed in the other panels.
The central panels (c,d) refer to the case of a pumping intensity just above the threshold. 
In this regime, the condensate density is very small and the real part of the Bogoliubov mode frequency reduces to the single-particle one. 
On the other hand, the characteristic damping rate of density fluctuations (the gapped mode at $q=0$) and of the high-momentum modes is suppressed by a critical slowing down phenomenon~\cite{nonresonant}.
As a result, the low energy modes around $q\simeq -k_c$ become dynamically unstable. 
Eventually, this mechanism leads to the disappearence of the original condensate and the formation of another -stable- condensate in a lower momentum state. 

\begin{figure}[htpb]
\begin{center}
\includegraphics[width=0.95\columnwidth,angle=0,clip]{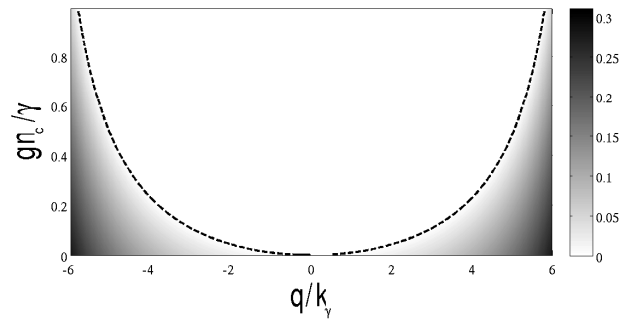}  
\end{center}
\caption{Growth rate (in units of $\gamma$) of maximally unstable mode as a function of condensate momentum $k_c$ and density $n_c$. The dashed line indicates the boundary of the stability region. Same system parameters as in Fig.\ref{fig:lindisp}(c-f). }
\label{fig:stab}
\end{figure}

The right panels (e,f) refer to the case of a stronger pumping well above the threshold. In this case, the damping rate of all modes other than the Goldstone one is comparable to the polariton lifetime $\gamma$. In this case, the frequency-dependence of pumping is not able to destabilize the moving condensate.
The stability domain as a function of the momentum $k_c$ and the density $n_c$ of the condensate is summarized in Fig.\ref{fig:stab}: as the density is increased, stable condensates survive up to larger momenta.

It is of crucial importance to note that this instability mechanism has no direct counterpart in standard equilibrium superfluids and is effective even in the absence of any defect potential.
 Of course, the present theory is directly applicable only to infinite, homogeneous condensates or in multiply connected ring-shaped geometries. 
In finite systems, assessing the stability of moving condensates would be made harder by the flow of condensate polaritons outside the pump regions~\cite{noi-shape}.

In the presence of weak defects,  the frictionless flow of equilibrium condensates is limited to flow speeds below the Landau critical velocity
$v_c=\min_k\left\{{\textrm{Re}[\omega_{Bog}^o(k)]}/{k}\right\}$,
where $\omega_{Bog}^o(k)$ is the real part of the Bogoliubov dispersion for a condensate at rest.
For dilute systems, $v_c$ coincides with the sound velocity $c_s=\sqrt{g n_c /m}$~\cite{SSLP-book}.
As a consequence of the diffusive character of the Goldstone mode, a na\"ive application of this definition to the Bogoliubov dispersion of non-equilibrium condensates anticipated in~\cite{nonresonant,keeling,Goldstone} gives a vanishing prediction for the critical velocity, $v_c=0$: Bogoliubov waves are emitted in a moving condensate hitting a (weak) defect at any  value of the condensate speed. 

\begin{figure}[htpb]
\begin{center}
\includegraphics[width=\columnwidth,angle=0,clip]{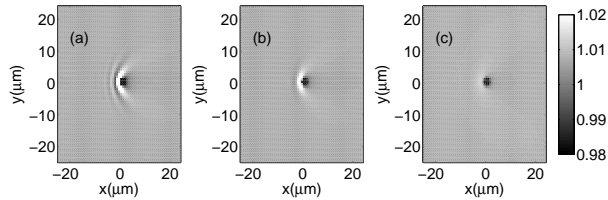}  
\end{center}
\caption{Density perturbation created in a moving condensate by a stationary weak defect for three different values of the condensate velocity $v/c_s=1.5,1,0.4$ across the critical value for superfluidity. Parameters: $n_c g/\gamma= n_c r/\gamma=1$, $\Omega_K/\gamma=50$.}
\label{fig:perturb_1}
\end{figure}

A complete picture of non-equilibrium condensates requires taking into account the non-trivial dynamics of the imaginary part of the Bogoliubov dispersion.
Numerical plots of the density perturbation induced by a single weak stationary defect in a (dynamically stable) moving condensate can be calculated from the generalized GPE \eq{CGLE} in the presence of a suitable defect potential $V_{ext}(\rr)$ and are shown in Fig.\ref{fig:perturb_1} for different values of the condensate velocity.
In contrast to the predictions of the na\"ive Landau criterion stated above, the induced perturbation closely resembles the one induced in an equilibrium condensate described by the standard GPE~\cite{Cerenkov}:
at high speeds [panel (a)], the defect creates a series of parabolic fringes that propagate away from the defect; at low speeds [panel (c)], the propagating fringes are replaced by a localized perturbation in the vicinity of the defect.
Remarkably, the characteristic speed at which the fringes disappear is of the order of the equilibrium sound speed $c_s=\sqrt{g n_c /m}$, but does not correspond to any noticeable feature in the real part of the Bogoliubov dispersion of Fig.\ref{fig:lindisp}(a,c,e).

\begin{figure}[htpb]
\begin{center}
\includegraphics[width=1.\columnwidth,angle=0,clip]{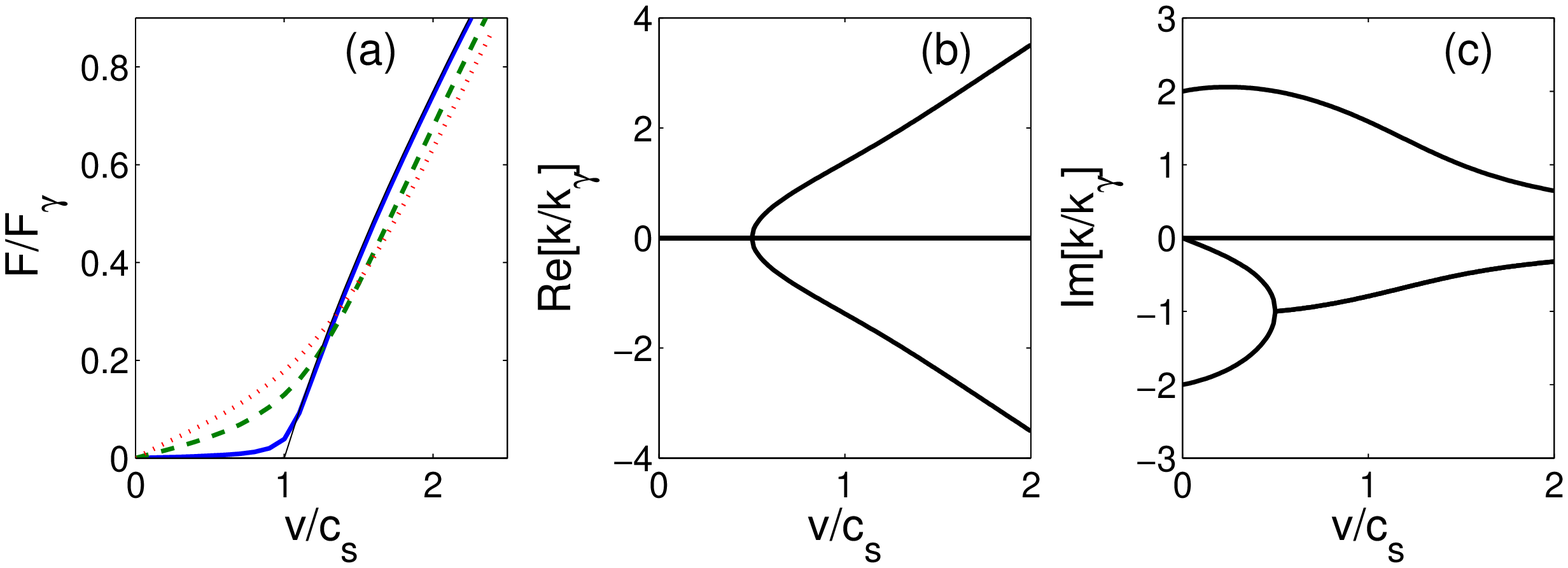}  
\end{center}
\caption{Panel (a): Force exerted on a weak stationary defect by a moving condensate as a function of the condensate velocity for different values of $r n_c /\gamma=0.1, 1, 2$ (blue solid, green dashed, red dotted lines), in units of $F_\gamma=\hbar^2 k_\gamma^3/m$. For comparison, the same force on an equilibrium condensate is shown as a thin solid black line. Panel (b,c): Real and imaginary parts of the complex wavevector $\tilde{k}$ of the Bogoliubov emission in the upstream direction as a function of the condensate velocity, for $rn_c /\gamma=gn_c /\gamma=1$. The other parameters are the same as in Fig.~\ref{fig:perturb_1}.}
\label{fig:force}
\end{figure}
 
As it was pointed out in a different context in~\cite{tait}, the emission pattern by a localized monochromatic source is better understood in terms of the (complex) wavevector of the emitted wave at the given (real) frequency. As one can see in Fig.~\ref{fig:force} (b), the wavevector $\tilde{k}$ of the Bogoliubov wave emitted at zero frequency by the static defect starts having a finite real part only after a branching point: extended oscillations in the density are observed as soon as the real part of $\tilde{k}$ exceeds the imaginary part.

This behavior is reflected in the drag force $F$ exerted by the moving condensate onto the defect as a function of the condensate velocity [Fig.\ref{fig:force}(a)]. 
In terms of the perturbed density profile $n(\rr)$, the drag force is given by
$F=-\int\!d^2\rr\,n(\rr)\,\nabla_\rr V_{def}(\rr)$~\cite{grisha}.
As a consequence of the finite lifetime of the Bogoliubov modes, the drag force has a non-vanishing value at all $v$. In addition, the drag force shows a marked threshold at a velocity value that closely corresponds to the onset of fringes in the density profile: the lower the value of the $n_cr/\gamma$ parameter, the sharper the threshold. On the other hand, the role of $\Omega_K$ on this effect is minor.
Even though no experiment has so far investigated the superfluid properties of polariton condensates under non-resonant pumping, it is likely that this generalized form of the Landau criterion provides at least a partial explaination of the recent experimental observation of superfluidity in a OPO regime~\cite{TOPO}.

Supersonic flows in equilibrium condensates in ring-shaped geometries are generally strongly sensitive to the presence of defects: as soon as nodes appear in the condensate wavefunction, the topological stability of the supercurrent state is broken, which leads to a rapid slow down of the condensate motion.
The finite damping rate of excitations in polariton condensates introduces a substantial modification to this picture: the perturbation created by each defect is not able to propagate on long distances, but rather remains localized in space on a length scale inversely proportional to $\textrm{Im}[\tilde{k}]$. As a result, we expect it is much harder to break the topological stability of the supercurrents.

\begin{figure}[htbp]
\begin{center}
\includegraphics[width=\columnwidth,angle=0,clip]{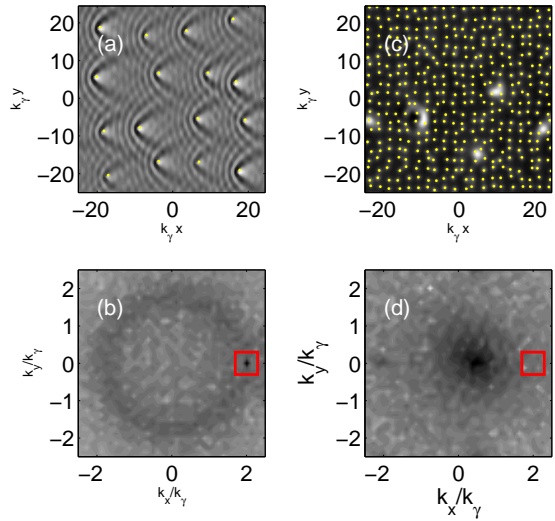}
\end{center}
\caption{Real (linear color scale) and momentum space (logarithmic color scale) densities of a non-equilibrium condensate after a temporal evolution of $\gamma t=300$ for an initial momentum $k/k_\gamma=2$ and two different densities of defects. The dots in the real space panels indicate the position of the defects. The square in the momentum space panels indicates the initial momentum of the condensate. Parameters are the same as in Fig.~\ref{fig:force}, except for $\Omega_K/\gamma=10$.}
\label{fig:weakdefect}
\end{figure}

This physical picture is confirmed by numerical simulations of the time evolution under the generalized GPE \eq{CGLE} starting from an initial condition with a supersonic flow. Periodic boundary conditions are assumed.
For a low defect density, the supercurrent state is maintained for very long times  with no sign of decay.  The characteristic Cerenkov-like density patterns in the vicinity of each defect are spatially separated and do not interfere [Fig.\ref{fig:weakdefect}(a)]. The momentum distribution in $k$-space [Fig.\ref{fig:weakdefect}(b)] shows the condensate peak right at the initial momentum state and a much fainter resonant Rayleigh scattering ring~\cite{iac-superfl}. 
The situation is different for a higher density of defects. In this case, the real space density perturbations created by the different defects substantially overlap with each other [Fig.\ref{fig:weakdefect}(c)]. As a result, interference effects are more likely to create nodes in the condensate wavefunction and therefore to trigger dissipation of the supercurrent. The occurrence of such a process is apparent in Fig.\ref{fig:weakdefect}(d) as a much reduced late-time value of the condensate momentum.

This fact illustrates another important difference with respect to standard, equilibrium condensates: in that case, the dissipation of a supercurrent leads to a significant heating and reduction of the condensate fraction. In the present case, the condensate turns out to be almost rigidly transferred to a lower momentum state; the (coherent) momentum broadening that is visible in Fig.\ref{fig:weakdefect}(d) is an effect of the defects. Remarkably, the frequency dependence of pumping in the generalized CGLE \eq{CGLE} has a crucial role in concentrating the population in low-energy modes and preventing the momentum distribution from spreading all over the $k$-space. 

In conclusion, we have investigated how the non-equilibrium nature of exciton-polariton condensates affects their superfluidity properties. A mean-field model based on a generalized Gross-Pitaevskii equation including a frequency-dependent pumping is developed. Different aspects of superfluidity have been considered. A non-equilibrium version of the Landau critical speed for the onset of drag force is formulated and metastability of supercurrents is shown to persist even at speeds well above the critical speed. 

We are indebted to Vincenzo Savona, Cristiano Ciuti and Davide Sarchi for continuous enlightening exchanges. Stimulating discussions with J. Keeling, C. Menotti, F. Piazza, D. Sanvitto, A. Smerzi are acknowledged.


\begin{thebibliography}{99}

\bibitem{SSLP-book} L.P. Pitaevskii and S. Stringari, \textsl{Bose-Einstein
Condensation}, Clarendon Press Oxford (2003).

\bibitem{leggett} A. J. Leggett, Rev. Mod. Phys. {\bf 71}, S318 (1999).



\bibitem{huang} K. Huang, {\em Statistical Mechanics} (New York, John Wiley \& Sons, 1963).


\bibitem{BEC-magnons} O. Demokritov, {\em et al.}, 
Nature {\bf 443}, 430 (2006).	

\bibitem{kasprzak} M. Richard {\em et al.}, Phys. Rev. B {\bf 72},
  201301(R) (2005); J. Kasprzak {\em et al.}, Nature {\bf 443}, 409 (2006).

\bibitem{maxime-small} M. Richard {\em et al.}, Phys. Rev. Lett. {\bf 94}, 
187401 (2005).

\bibitem{iac-superfl} I.Carusotto, C.Ciuti, Phys. Rev. Lett. {\bf 93}, 166401 (2004).

\bibitem{vortexprobe} M. Wouters and V. Savona, arXiv:0904.2966.

\bibitem{amo-superfl}  A. Amo, {\em et al.}
Nature Phys. {\bf 5}, 805 (2009).

\bibitem{TOPO} A. Amo, {\em et al.}, 
Nature {\bf 457}, 291 (2009).


\bibitem{vortexMadrid} D. Sanvitto, {\em et al.}
arXiv:0907.2371. 









\bibitem{Goldstone} M. Wouters, I. Carusotto, Phys.Rev. A {\bf 76}, 043807 (2007).


\bibitem{nonresonant} M. Wouters and I. Carusotto, Phys. Rev. Lett. {\bf 99}, 140402 (2007).

\bibitem{keeling} J. Keeling and N. G. Berloff, Phys. Rev. Lett. {\bf 100}, 250401 (2008).

\bibitem{noi-shape} M. Wouters, I. Carusotto, and C. Ciuti, Phys. Rev. B {\bf 77}, 115340 (2008) 

\bibitem{synchro} A. Baas, {\em et al.}, Phys. Rev. Lett.{\bf  100}, 170401 (2008); M. Wouters,
Phys. Rev. B {\bf 77}, 121302 (2008); P. R. Eastham, Phys. Rev. B {\bf 78}, 035319(R) (2008).

\bibitem{vortices} K. G. Lagoudakis {\em et al.}, Nature Physics {\bf 4}, 706 (2008).

\bibitem{kinetic} D. Porras, C. Ciuti, J. J. Baumberg, C. Tejedor, Phys. Rev. B {\bf 66}, 085304 (2002).

\bibitem{fluctuations} M. Wouters, V. Savona, Phys.Rev.B {\bf 79}, 165302 (2009).



\bibitem{tait} W. C. Tait, Phys. Rev. B {\bf 5}, 649 (1972).

\bibitem{grisha}  G. E. Astrakharchik and L. P. Pitaevskii, Phys. Rev. A {\bf 70}, 013608 (2004).

\bibitem{Cerenkov} I. Carusotto, S. X. Hu, L. A. Collins, and A. Smerzi
Phys. Rev. Lett. {\bf 97}, 260403 (2006).

\end{thebibliography}
\end{document}